\documentclass[electronic]{vgtc}
\graphicspath{{figures/}{pictures/}{images/}{./}}
\usepackage{times}

\usepackage{mathptmx}
\usepackage{booktabs}
\usepackage{tabularx}
\usepackage{enumitem}
\usepackage{microtype}
\usepackage{amsmath}
\usepackage{xspace}

\onlineid{1085}
\vgtccategory{Research}
\vgtcinsertpkg
\nocopyrightspace

\newcommand{\system}{\textsc{EvidenceLens}\xspace}

\title{\system: A Claim-Evidence Matrix for Auditing Financial Question Answering}

\author{Fengchen Gu$^1$, Xiaotian Ren$^1$, Zhengyong Jiang$^1$, Zhilu Zhang$^1$, \'{A}ngel F. Garc\'{i}a-Fern\'{a}ndez$^2$, \\ Angelos Stefanidis$^1$, Mian Zhou$^1$, Huakang Li*$^1$, Jionglong Su*$^1$}
\affiliation{\scriptsize $^1$ School of AI and Advanced Computing, XJTLU Entrepreneur College (Taicang), Xi’an Jiaotong-Liverpool University, Suzhou, Jiangsu, China \\ $^2$ ETSI de Telecomunicaci\'{o}n, Universidad Polit\'{e}cnica de Madrid, Madrid, Spain \\ IEEE VIS 2026 conditionally accepted version}

\abstract{
Large language models are increasingly used to answer questions over annual reports, earnings decks, and analyst notes, yet their outputs remain difficult to verify in high-stakes financial workflows. A fluent answer can blend directly grounded statements, weak synthesis, and unsupported claims across narrative text, tables, and charts. We present \system, a visual analytics prototype that treats financial question answering as a claim-evidence alignment problem. The system decomposes an answer into atomic claims, summarizes support composition and confidence, support gaps, and coordinates claim-level inspection with source passages, table cells, and chart regions. Its core visual representation is a multimodal claim-evidence matrix that makes coverage, contradiction, and modality imbalance immediately visible. To support reproducibility, we also specify a JSON-based artifact schema, a lightweight multimodal alignment pipeline, and a deterministic review-priority ranking that maps backend signals into an auditable visual structure. Through representative report-auditing scenarios, we show how \system helps analysts distinguish grounded claims from overconfident synthesis that conventional chat interfaces flatten.
}


\begin{document}
\firstsection{Introduction}
\maketitle

Financial analysts increasingly use LLM-based copilots to summarize earnings materials, answer due-diligence questions, and accelerate report review. The appeal is obvious: a single prompt can condense dozens of pages of management discussion, tables, and exhibits into a short answer. The risk is equally obvious. A polished response can mix directly grounded statements with speculative synthesis, while conventional chat interfaces provide only coarse citation links or no provenance at all. In finance, where users must defend decisions and audit the basis of claims, ``looks plausible'' is not a sufficient criterion.

Recent work in finance QA has made progress on reasoning over hybrid report content, including tables and surrounding text \cite{finqa,tatqa}. Chart question answering likewise highlights the difficulty of aligning natural-language questions with visual marks and logical operations \cite{chartqa}. In visualization, researchers have argued that communicative understanding depends on explicit fact--evidence reasoning \cite{vaidyadasgupta2020}, and that automated insight systems need clearer conceptual structure \cite{law2020}. Yet these perspectives have rarely been combined in an interface for auditing LLM answers in a financial setting, where evidence is simultaneously textual, tabular, and visual.

Recent visualization research has also expanded rapidly around AI-mediated chart understanding and oversight. Studies examine whether VLMs truly understand charts and how stable their interpretations remain across layouts, prompts, and deceptive designs \cite{mukhopadhyay2024unraveling,wang2025aligned,lo2025misleading,chartdeception2025}. Other work explores whether visualization itself helps AI reason over data \cite{vizhelpsai2025}, how multimodal models can be adapted to visualization understanding \cite{simvecvis2025}, and how LLMs participate in design feedback, storytelling, and retargeting workflows \cite{ahn2025chatgpt,reflection2025,llmretargeting}. Complementary VIS papers address integrity, recourse, uncertainty, and explanation interfaces \cite{visualintegrity2025,groundedchart2025,bnnvis2025,revise2025,xaireverse2025,logicdecisionaid2025}. On the finance side, newer multimodal benchmarks and retrieval pipelines increasingly target long reports containing prose, tables, and figures \cite{finsage2025,multifinrag2025,multifinben2025,wikimixqa2025,aida2025layout}. However, these lines still leave a gap between answer generation and analyst verification.

We argue that financial QA should be treated as a \emph{claim-evidence alignment} problem rather than a text-generation problem alone. Users need to inspect not only \emph{where} an answer is cited, but also \emph{which claim} is supported, \emph{by what modality}, and \emph{whether conflicting evidence exists elsewhere}. This framing motivates \system, a compact visual analytics prototype for auditing multimodal financial answers.

\system makes four concise contributions. First, it introduces a claim-centered provenance representation for financial QA over text, tables, and charts. Second, it contributes a multimodal claim-evidence matrix that exposes support coverage, modality imbalance, and contradiction at analyst-audit speed. Third, it specifies a lightweight but reproducible alignment and review-priority pipeline with explicit weights, thresholds, and artifact schemas. Fourth, it demonstrates the design through representative report-auditing scenarios and distills lessons for trustworthy financial AI interfaces. This focus aligns with the system/tool and exploratory application categories emphasized by the VIS short-paper format \cite{pygwalker}.

\section{From Financial QA to Claim-Evidence Alignment}
A typical financial answer contains several distinct statements: a factual observation (``operating margin increased''), an explanatory inference (``the increase came from procurement savings''), and a broader judgment (``one-off items were not material''). Treating the whole answer as a single unit hides precisely the distinctions analysts care about. \system therefore models an answer as a set of atomic claims $C=\{c_1,\dots,c_n\}$, each linked to zero or more evidence items $E=\{e_1,\dots,e_m\}$.

An evidence item may be a text span, a table cell or row, or a chart mark or time interval. We normalize each item as $e_j=(m_j,u_j,t_j,a_j)$, where $m_j \in \{\text{text},\text{table},\text{chart}\}$ is the modality, $u_j$ is a modality-specific anchor, $t_j$ is an indexable surface description, and $a_j$ stores typed attributes. For tables, the anchor records a cell together with row and column header paths; for charts, it records a mark group or x-interval with derived attributes such as slope, extrema, or delta. Claims are also typed as value, comparison, trend, or explanation so that later alignment can use modality-aware checks. We further normalize each claim as $c_i=(y_i,o_i,v_i,q_i)$, where $y_i$ is the claim type, $o_i$ is an operator template (increase, decrease, higher-than, attributable-to, not-material), $v_i$ is an optional numeric payload, and $q_i$ is a normalized text gloss. This compact representation makes the downstream checks interpretable: a trend claim expects directional evidence, a value claim expects numeric agreement, and an explanation claim expects at least one semantically aligned rationale span plus, when available, corroborating quantitative context.

\begin{figure*}[t]
  \centering
  \includegraphics[width=\linewidth]{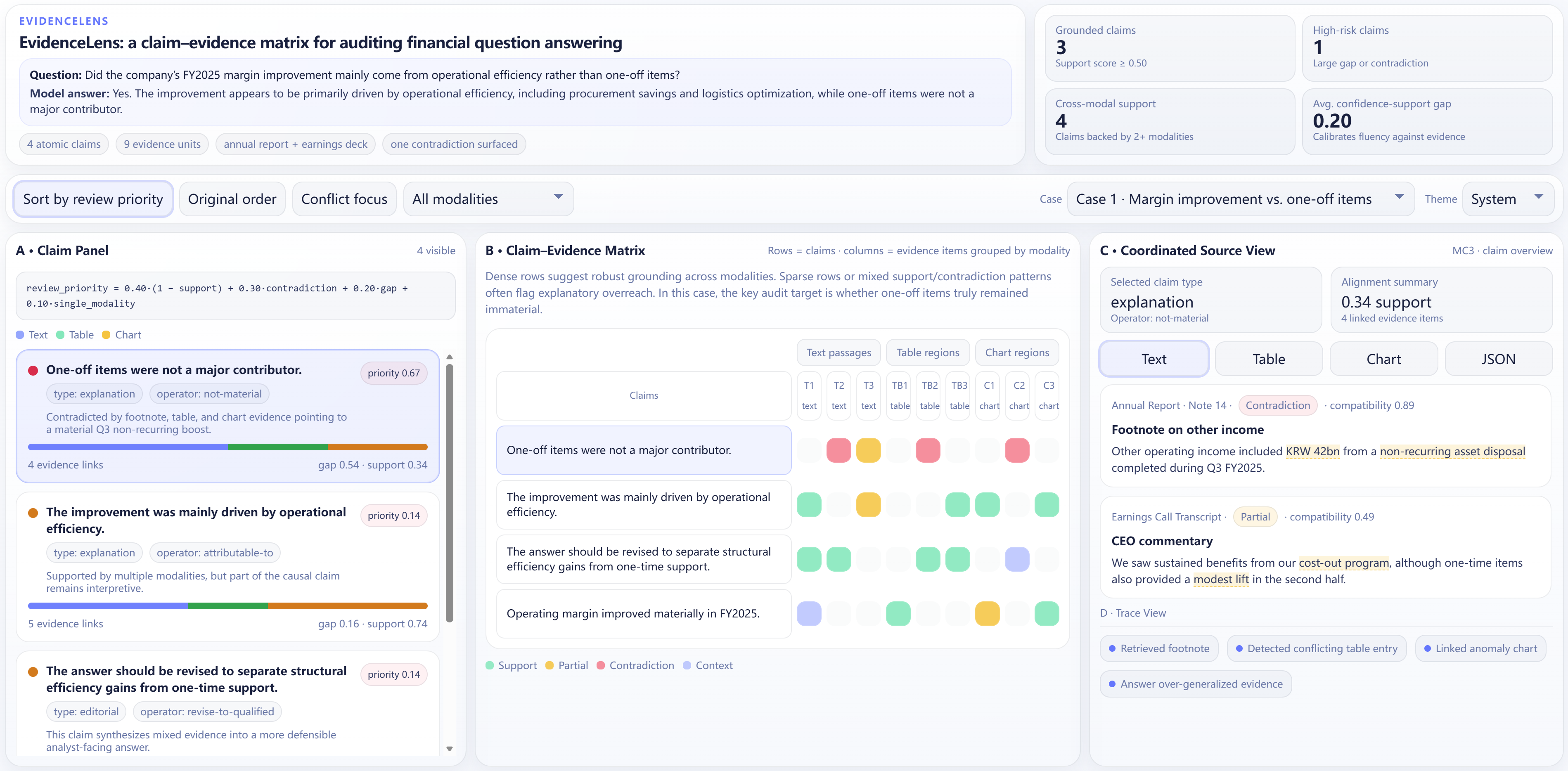}
  \caption{\system converts a generated financial answer into an auditable claim-evidence representation. \textbf{A} The Claim Panel decomposes the answer into atomic claims and summarizes support composition, confidence--support gaps, and review priority. \textbf{B} The central Claim-Evidence Matrix groups evidence columns by modality and source order, making sparse support, cross-modal corroboration, and contradiction visible at a glance. \textbf{C} The Coordinated Source View preserves local context through synchronized highlights in text, tables, and charts. \textbf{D} The Trace View summarizes the evidence path used to assemble the answer.}
  \label{fig:overview}
\end{figure*}

For each claim-evidence pair, the prototype computes a lightweight compatibility score
\[
\begin{aligned}
 s_{ij} ={}& w_{\mathrm{sem}}\phi_{\mathrm{sem}}(c_i,e_j) + w_{\mathrm{num}}\phi_{\mathrm{num}}(c_i,e_j) \\
 &+ w_{\mathrm{str}}\phi_{\mathrm{str}}(c_i,e_j) + w_{\mathrm{loc}}\phi_{\mathrm{loc}}(c_i,e_j),
\end{aligned}
\]
where $\phi_{\mathrm{sem}}$ measures embedding similarity between the claim and evidence description, $\phi_{\mathrm{num}}$ checks agreement of numbers, directions, or comparative operators, $\phi_{\mathrm{str}}$ measures compatibility between claim type and evidence anchor, and $\phi_{\mathrm{loc}}$ rewards local corroboration from nearby report regions. We label a link as \emph{support} when $s_{ij}\geq \tau_s$ and no polarity conflict is detected, \emph{partial support} when $\tau_p\leq s_{ij}<\tau_s$, \emph{contradiction} when numeric or directional mismatch is detected, and \emph{context} otherwise. This representation is intentionally backend-agnostic: stronger production systems can replace our simple scoring without changing the visualization. We operationalize contradiction with a polarity test $\kappa_{ij}$ that fires when the sign or comparative direction extracted from $c_i$ conflicts with that of $e_j$, or when a numeric value falls outside an allowed tolerance band. For chart evidence, the detector uses derived attributes in $a_j$ such as local slope, extrema, or one-step deltas; for tables, it compares parsed values after unit normalization; for text, it relies on lexical negation and directional verbs. Although approximate, this detector is sufficient to separate weak support from active disagreement, which is the distinction analysts most urgently need.

This data model is driven by three design goals. \textbf{G1: expose answer structure at the claim level.} Users must judge specific statements, not an answer blob. \textbf{G2: unify heterogeneous evidence.} Financial audit tasks routinely require movement between prose, numerical tables, and charts. \textbf{G3: foreground weak support.} In practice, analysts spend most of their time identifying unsupported or contradicted statements, so the interface should prioritize those cases visually.

\section{Visual Design of \system}
\cref{fig:overview} shows the interface. Its center of gravity is not the chat transcript but the \emph{structure of evidence} behind the answer.

\paragraph{Claim Panel.}
The left panel decomposes an answer into claim cards. Each card shows a status glyph, the claim text, and a stacked support bar encoding the relative contribution of text, table, and chart evidence. We also visualize a \emph{confidence--support gap}: if model confidence is high but aligned support is weak, the claim is prioritized for review. This design replaces citation footnotes with a fast triage view. A user can immediately distinguish a claim backed by multiple modalities from one supported by a single soft text span.

\paragraph{Evidence-Alignment Matrix.}
The central matrix is the primary visualization contribution. Rows correspond to claims; columns correspond to evidence items grouped by modality and source order. Each cell encodes the relation between a claim and an evidence item using support type. Dense rows suggest broad corroboration; sparse rows suggest thin grounding; mixed green/red patterns expose internal conflict. Compared with sequential citation lists, the matrix supports side-by-side comparison across claims and makes support coverage visually legible. It also surfaces modality imbalance, such as an explanatory claim justified only by prose without quantitative backing. In our implementation, columns are ordered first by modality and then by document position so that adjacent columns often share local context. This stabilizes the visual layout across repeated questions on the same report and makes it easier to compare two claims against the same evidence neighborhood.

\paragraph{Coordinated Source View.}
Selecting a row or cell updates synchronized source tabs for text, tables, and charts. Text evidence is highlighted in situ so surrounding sentences remain visible. Table evidence emphasizes both the selected cell and its headers, preserving semantic context. Chart evidence highlights the relevant visual region---for example, a quarter-to-quarter rise or a single-quarter anomaly---rather than merely citing an image. This design respects the fact that chart evidence is visual and relational, not simply textual. Importantly, selecting a matrix cell preserves bidirectional linkage: the source view reveals local context, while the originating row and column remain highlighted, reducing disorientation during rapid triage.

\paragraph{Trace View.}
A compact trace strip summarizes the order in which evidence items contributed to the answer. Although simple, this view is valuable when an answer emerges from multiple retrieval and reasoning steps; it helps users see whether the system over-relied on one passage or whether support emerged from several independent sources.

\paragraph{Interaction.}
The prototype supports three forms of triage. \emph{Risk sorting} ranks claims by confidence--support gap and contradiction count so that review starts with the most fragile statements. \emph{Conflict focus} filters the matrix to rows containing both support and contradiction, a frequent pattern when management narrative and footnotes diverge. \emph{Modality filtering} isolates text-only, table-only, chart-only, or cross-modal claims. Together these interactions preserve an overview-first workflow while keeping the evidence matrix as the central analysis surface. We deliberately avoid heavy natural-language explanation overlays in the overview because they compete with structure; instead, explanatory notes are attached to details-on-demand so that the analyst can maintain a global picture of coverage before reading prose.

\section{Prototype Instantiation}
Our current prototype is implemented as a lightweight single-page interface and uses a deliberately simple backend to foreground visualization concerns. Given a question and draft answer, the system first segments the answer into claim candidates using sentence boundaries, discourse connectors, and a small operator lexicon for explanation, comparison, and trend language. It then aligns each claim with evidence units extracted from narrative passages, table regions, and chart intervals. Alignment can be supplied manually, but our prototype also supports a semi-automatic workflow in which an LLM proposes claims and rationale spans and a small ruleset maps the rationale to source regions. Tables are linearized into cell-plus-header paths, while chart regions are paired with short derived captions such as ``other income spikes in Q3'' or ``gross margin rises from Q1 to Q4.'' This hybrid setup is sufficient for exploratory design because it exposes the same uncertainty patterns a stronger production backend would still need to communicate.

To make the prototype reproducible, each case is stored as a single JSON instance with three top-level arrays: \texttt{claims}, \texttt{evidence}, and \texttt{links}. A claim record contains an identifier, normalized type, surface text, and model confidence. An evidence record stores modality, source location, anchor metadata, and a short textual description used by the alignment stage. A link record stores the pair $(c_i,e_j)$ together with its score and discrete label. Once this file is fixed, the claim list, matrix, and claim-summary CSV are rebuilt deterministically. The accompanying artifact package includes two cases, a JSON schema, a configuration file with default parameters, and a small script that reconstructs the matrix and risk ranking used by the interface. This gives the paper a reproducible boundary: backend components may vary, but the same claim/evidence instance always yields the same visual state and review ordering.

To prioritize review, we aggregate per-modality support instead of simply counting links. Let $M_i$ be the set of modalities that provide at least one non-context link for claim $c_i$. The prototype computes claim support as
\[
 \mathrm{Supp}(c_i)=\frac{1}{|M_i|}\sum_{m\in M_i}\max_{e_j:m_j=m} s_{ij},
\]
which prevents many weak text spans from overwhelming a single strong quantitative item. We additionally flag explanatory claims as weak when $y_i=\text{explanation}$ and $\mathrm{Supp}(c_i)<\tau_e$, because purely semantic support is often insufficient for analyst-facing answers. Each claim then receives a review priority
\[
\begin{aligned}
 r_i ={}& \alpha \, \max\bigl(0,\mathrm{conf}(c_i)-\mathrm{Supp}(c_i)\bigr) + \beta\,\mathbf{1}[\exists\,\text{contradiction}] \\
 &+ \gamma\,\mathbf{1}[|M_i|=1] + \delta\,\mathbf{1}[y_i=\text{explanation} \land \mathrm{Supp}(c_i)<\tau_e].
\end{aligned}
\]
In the released default configuration we use $(w_{\mathrm{sem}},w_{\mathrm{num}},w_{\mathrm{str}},w_{\mathrm{loc}})=(0.35,0.30,0.20,0.15)$, $(\tau_s,\tau_p,\tau_e)=(0.72,0.55,0.65)$, and $(\alpha,\beta,\gamma,\delta)=(0.5,0.3,0.1,0.1)$. While intentionally lightweight, this scoring operationalizes the key idea that analysts should review claims not only when support is missing, but also when the system sounds overly certain or leans on a single modality. In practice, this pushes explanatory and judgmental claims above straightforward factual claims, which matches real review priorities in finance.

\begin{table*}[t]
\caption{Representative financial QA failure modes that motivated the design.}
\label{tab:failuremodes}
\centering
\resizebox{\linewidth}{!}{%
\begin{tabular}{@{}lll@{}}
\toprule
Failure mode & Visual symptom in \system & Review action \\
\midrule
Unsupported explanatory leap & Sparse row with support concentrated in prose & Check whether numbers or charts corroborate the explanation \\
Hidden contradiction in fine print & Mixed support/contradiction cells in one row & Inspect footnotes, note tables, and anomaly charts \\
Overconfident synthesis & Large confidence--support gap despite clean language & Revise answer to a qualified statement \\
\bottomrule
\end{tabular}%
}
\end{table*}

For reproducibility, the matrix is constructed from a fixed row order over claims and a fixed column order over evidence units. The released reconstruction script writes two deterministic artifacts for each case: a claim$\times$evidence CSV and a claim-summary CSV containing support scores, modality coverage, contradiction flags, and risk scores. This separation is useful for future work because stronger retrieval or chart-parsing modules can be swapped in while preserving the same downstream visual encoding and evaluation logic.

\section{Scenario Walkthroughs}
We illustrate \system using two representative questions over an annual report and an earnings deck.

\paragraph{Case 1: Margin improvement.}
The question asks whether FY2025 margin improvement mainly came from operational efficiency rather than one-off items. A baseline chat answer reads as a confident ``yes.'' In \system, the first claim (operating margin improved) appears strongly grounded by both a table row and a chart trend. The second claim (improvement mainly came from operational efficiency) is supported, but more unevenly: management discussion text and a cost-bridge table align, while some of the explanation remains inferential. The third claim (one-off items were not a major contributor) stands out as problematic. In the matrix, its row is sparse and mixed, with contradiction from a footnote on asset disposal, a table showing a jump in other operating income, and a chart spike in Q3. The key benefit is not simply finding a wrong citation; it is seeing that the answer is \emph{partially grounded and partially contradicted}. This distinction is difficult to perceive in a conventional chat transcript.

\paragraph{Case 2: Guidance revision.}
A second scenario asks whether a guidance cut reflects demand weakness or temporary channel inventory normalization. Here the answer includes both textual management commentary and a line chart on shipment trends. \system helps disentangle two patterns. First, some claims are well covered by narrative passages but not by numbers; second, the chart supports a short-term inflection but not a broader demand-collapse interpretation. The matrix makes this difference explicit: support concentrates in contextual text cells, while quantitative evidence remains narrow. This helps an analyst revise the answer from a categorical claim to a qualified one.

Across both cases, we observed three recurrent failure modes that the visualization makes legible: unsupported explanatory leaps, contradictory fine print that is hidden behind broad summaries, and overconfidence when support is concentrated in a single modality. \cref{tab:failuremodes} summarizes how these failures appear in the interface and how they redirect analyst attention. The same answer may therefore contain claims that should be accepted, claims that should be qualified, and claims that should be rejected outright. That granularity is precisely what conventional chat interfaces flatten.

\section{Analyst Workflow and Design Rationale}
\system is designed around a realistic audit loop rather than open-ended exploration. In practice, an analyst begins with a draft answer, scans the claim list for risky statements, then drills into one claim at a time to decide whether it should be accepted, qualified, or rewritten. The interface therefore uses an overview-first progression with increasingly grounded details: claims first, then cross-claim evidence structure, then original source context. This sequence is important. If the interface begins with raw passages and tables, users are forced back into manual document search. If it begins and ends with chat text, they never see the shape of the supporting argument.

We also considered alternative representations for claim-evidence linkage. A node-link graph is expressive, but it becomes visually noisy as soon as multiple evidence items connect to several claims. A threaded citation list preserves reading order, but makes side-by-side comparison across claims difficult. The matrix form is more austere, yet it is better matched to the analyst task: it supports rapid comparison, reveals coverage patterns through density, and naturally accommodates modality grouping. In other words, the matrix favors \emph{auditability} over narrative flourish.

A second design choice is to treat charts as first-class evidence rather than screenshots attached to citations. Financial reasoning often depends on visual patterns such as slope changes, single-quarter spikes, or the shape of a bridge chart. These semantics do not map cleanly onto sentence spans. By linking claims to chart regions, the interface allows users to inspect whether a verbal statement genuinely follows from the depicted trend or whether the answer overstates what the chart shows. This distinction matters in finance because visual summaries in decks often carry more interpretive weight than the surrounding prose.

Finally, \system makes calibration explicit. Conventional QA systems often present answer confidence, but confidence alone is hard to interpret without a view of support quality. The confidence--support gap gives users a more actionable signal: a claim can be linguistically fluent and internally confident while still resting on thin or conflicting evidence. We found this signal particularly useful for explanatory and causal claims, which are precisely the ones analysts most need to interrogate.

\section{Toward Evaluation}
Although this paper focuses on design, the prototype suggests a clear evaluation path. A compact comparative study can pair \system with a citation-based baseline chat interface and assign participants three common review tasks: identify unsupported claims, locate contradictory evidence, and decide whether an answer should be accepted, revised, or rejected. The central hypotheses are that claim-level provenance will improve unsupported-claim detection, reduce verification time for mixed-modality questions, and better calibrate user confidence. Because the matrix and risk ranking are deterministically reconstructed from fixed instances, the same cases can also support repeatable ablations over row ordering, modality grouping, or ranking weights without changing the underlying evidence.

A concrete study can use 12--16 report questions spanning value lookup, trend explanation, and conflict resolution. In addition to task time and final judgment accuracy, the interface makes it possible to measure unsupported-claim recall, contradiction-localization time, false-alarm rate, calibration gap between trust judgment and answer quality, and modality coverage (which tabs or evidence types users inspected before deciding). These measures are directly relevant to financial oversight settings, where speed matters but missed contradictions are costly. A useful ablation plan would compare four conditions: answer text only, answer plus citations, claim list without the matrix, and the full interface. This isolates whether the gain comes from claim decomposition alone or from the matrix's ability to expose cross-claim evidence structure. Because the prototype logs matrix clicks, tab switches, and claim reordering, it also supports process-level analysis of how users verify answers rather than only whether they succeed.

\section{Discussion}
Our prototype suggests that trustworthy financial QA depends as much on interface design as on model accuracy. First, \emph{claim decomposition} is essential: analysts evaluate statements, not paragraphs. Second, \emph{cross-modal provenance} matters because financial reasoning routinely spans prose, tables, and charts. Third, effective oversight requires making uncertainty and contradiction visually salient rather than burying them in expandable citations.

Beyond the specific prototype, we see two broader implications for visualization research. One is methodological: interfaces for AI-assisted analysis should expose not only retrieved sources but also the structure of argumentation across claims. The other is practical: finance provides a demanding testbed for provenance design because answers are short, documents are long, and evidence is heterogeneous. Designs that succeed here may generalize to policy, medicine, and scientific reporting.

There is also a broader systems implication. Many current financial copilots treat grounding as a retrieval problem and trust as a language problem. Our cases suggest a third layer is missing: a \emph{representation problem}. Even when relevant evidence is successfully retrieved, users still need a representation that lets them compare claims, notice imbalance, and preserve source context without losing the thread of the answer. Visualization is well positioned to supply this missing layer.

A further implication concerns authoring. Financial professionals rarely want a binary verdict of right or wrong; they want a revised answer that is precise enough to reuse in notes, slide titles, or internal memos. By making specific weak claims visible, \system supports this editorial workflow. In that sense, the interface is not only for error finding but also for answer refinement.

\system also has limitations. Claim extraction and evidence alignment are only as reliable as the upstream pipeline. Some financial judgments are legitimately synthetic and cannot be reduced to a single supporting item. We therefore treat the interface as an audit aid rather than a correctness oracle. Another limitation is scalability: a very long answer may require claim clustering or semantic folding to keep the matrix readable. These are promising directions for future work. A particularly important next step is scaling beyond short answers. For 20+ claims, the current matrix should be combined with claim clustering, row folding, or semantic facet filters so that analysts can move between summary-level audit and claim-level inspection without losing provenance.

\section{Conclusion}
\system reframes financial question answering as a visual auditing task. By decomposing answers into claims and aligning them with textual, tabular, and visual evidence, the interface reveals what is grounded, what is thinly supported, and what is contradicted. More broadly, it shows how a claim-evidence matrix can turn opaque multimodal QA into an auditable visual structure. We see this as a promising direction for financial AI tools that must be inspectable, defensible, and usable under time pressure.

\bibliographystyle{abbrv-doi}
\bibliography{paper}

@inproceedings{finqa,
  title = {FinQA: A Dataset of Numerical Reasoning over Financial Data},
  author = {Chen, Zhiyu and Chen, Wenhu and Smiley, Charese and Shah, Sameena and Borova, Iana and Langdon, Dylan and Moussa, Reema and Beane, Matt and Huang, Ting-Hao and Routledge, Bryan and Wang, William Yang},
  booktitle = {Proceedings of the 2021 Conference on Empirical Methods in Natural Language Processing},
  year = {2021},
  pages = {3697--3711},
  address = {Online and Punta Cana, Dominican Republic},
  publisher = {Association for Computational Linguistics},
  doi = {10.18653/v1/2021.emnlp-main.300},
  url = {https://aclanthology.org/2021.emnlp-main.300/}
}

@inproceedings{tatqa,
  title = {{TAT-QA}: A Question Answering Benchmark on a Hybrid of Tabular and Textual Content in Finance},
  author = {Zhu, Fengbin and Lei, Wenqiang and Huang, Youcheng and Wang, Chao and Zhang, Shuo and Lv, Jiancheng and Feng, Fuli and Chua, Tat-Seng},
  booktitle = {Proceedings of the 59th Annual Meeting of the Association for Computational Linguistics and the 11th International Joint Conference on Natural Language Processing (Volume 1: Long Papers)},
  year = {2021},
  pages = {3277--3287},
  address = {Online},
  publisher = {Association for Computational Linguistics},
  doi = {10.18653/v1/2021.acl-long.254},
  url = {https://aclanthology.org/2021.acl-long.254/}
}

@inproceedings{chartqa,
  title = {ChartQA: A Benchmark for Question Answering about Charts with Visual and Logical Reasoning},
  author = {Masry, Ahmed and Long, Do Xuan and Tan, Jia Qing and Joty, Shafiq and Hoque, Enamul},
  booktitle = {Findings of the Association for Computational Linguistics: ACL 2022},
  year = {2022},
  pages = {2263--2279},
  address = {Dublin, Ireland},
  publisher = {Association for Computational Linguistics},
  doi = {10.18653/v1/2022.findings-acl.177},
  url = {https://aclanthology.org/2022.findings-acl.177/}
}

@inproceedings{vaidyadasgupta2020,
  title = {Knowing What to Look For: A Fact-Evidence Reasoning Framework for Decoding Communicative Visualization},
  author = {Vaidya, Sahaj and Dasgupta, Aritra},
  booktitle = {2020 IEEE Visualization Conference (VIS)},
  year = {2020},
  pages = {231--235},
  publisher = {IEEE},
  doi = {10.1109/VIS47514.2020.00053}
}

@inproceedings{law2020,
  title = {Characterizing Automated Data Insights},
  author = {Law, Po-Ming and Endert, Alex and Stasko, John T.},
  booktitle = {2020 IEEE Visualization Conference (VIS)},
  year = {2020},
  pages = {171--175},
  publisher = {IEEE},
  doi = {10.1109/VIS47514.2020.00041}
}

@misc{pygwalker,
  title = {PyGWalker: On-the-fly Assistant for Exploratory Visual Data Analysis},
  author = {Yu, Yue and Shen, Leixian and Long, Fei and Qu, Huamin and Chen, Hao},
  year = {2024},
  eprint = {2406.11637},
  archivePrefix = {arXiv},
  primaryClass = {cs.HC},
  doi = {10.48550/arXiv.2406.11637},
  url = {https://arxiv.org/abs/2406.11637}
}

@inproceedings{mukhopadhyay2024unraveling,
  title = {Unraveling the Truth: Do {VLM}s really Understand Charts? A Deep Dive into Consistency and Robustness},
  author = {Mukhopadhyay, Srija and Qidwai, Adnan and Garimella, Aparna and Ramu, Pritika and Gupta, Vivek and Roth, Dan},
  booktitle = {Findings of the Association for Computational Linguistics: EMNLP 2024},
  year = {2024},
  pages = {16696--16717},
  address = {Miami, Florida, USA},
  publisher = {Association for Computational Linguistics},
  doi = {10.18653/v1/2024.findings-emnlp.973},
  url = {https://aclanthology.org/2024.findings-emnlp.973/}
}

@article{wang2025aligned,
  title = {How Aligned are Human Chart Takeaways and {LLM} Predictions? A Case Study on Bar Charts with Varying Layouts},
  author = {Wang, Huichen Will and Hoffswell, Jane and Thane, Sao Myat and Bursztyn, Victor S. and Xiong Bearfield, Cindy},
  journal = {IEEE Transactions on Visualization and Computer Graphics},
  year = {2025},
  volume = {31},
  number = {1},
  pages = {536--546},
  doi = {10.1109/TVCG.2024.3456378}
}

@article{lo2025misleading,
  title = {How Good (Or Bad) Are {LLM}s at Detecting Misleading Visualizations?},
  author = {Lo, Leo Yu-Ho and Qu, Huamin},
  journal = {IEEE Transactions on Visualization and Computer Graphics},
  year = {2025},
  volume = {31},
  number = {1},
  pages = {1116--1125},
  doi = {10.1109/TVCG.2024.3456333}
}

@inproceedings{llmretargeting,
  title = {Challenges \& Opportunities with {LLM}-Assisted Visualization Retargeting},
  author = {Snyder, Luke S. and Wang, Chenglong and Drucker, Steven M.},
  booktitle = {2025 IEEE Visualization and Visual Analytics (VIS)},
  year = {2025},
  pages = {141--145},
  publisher = {IEEE},
  doi = {10.1109/VIS60296.2025.00034}
}

@inproceedings{chartdeception2025,
  title = {The Perils of Chart Deception: How Misleading Visualizations Affect Vision-Language Models},
  author = {Mahbub, Ridwan and Islam, Mohammed Saidul and Laskar, Md Tahmid Rahman and Rahman, Mizanur and Nayeem, Mir Tafseer and Hoque, Enamul},
  booktitle = {2025 IEEE Visualization and Visual Analytics (VIS)},
  year = {2025},
  pages = {6--10},
  publisher = {IEEE},
  doi = {10.1109/VIS60296.2025.00006}
}

@inproceedings{vizhelpsai2025,
  title = {Does visualization help {AI} understand data?},
  author = {Li, Victoria R. and Sun, Johnathan L. and Wattenberg, Martin},
  booktitle = {2025 IEEE Visualization and Visual Analytics (VIS)},
  year = {2025},
  pages = {51--55},
  publisher = {IEEE},
  doi = {10.1109/VIS60296.2025.00016}
}

@inproceedings{visualintegrity2025,
  title = {Visual Integrity in the Age of {AI}: An Evaluation of {DLSS} and {DLAA} in Geospatial Visualization},
  author = {Beregovyi, Kindrat and Butkiewicz, Thomas},
  booktitle = {2025 IEEE Visualization and Visual Analytics (VIS)},
  year = {2025},
  pages = {291--295},
  publisher = {IEEE},
  doi = {10.1109/VIS60296.2025.00064}
}

@inproceedings{groundedchart2025,
  title = {Grounded Generation of Embellished Bar Chart Ensuring Chart Integrity},
  author = {Kim, Seon Gyeom and Choi, Jae Young and Lee, Yuseung and Chung, Jaeryung and Rossi, Ryan and Kil, Jihyung and Koh, Eunyee and Lee, Tak Yeon},
  booktitle = {2025 IEEE Visualization and Visual Analytics (VIS)},
  year = {2025},
  pages = {101--105},
  publisher = {IEEE},
  doi = {10.1109/VIS60296.2025.00026}
}

@inproceedings{simvecvis2025,
  title = {SimVecVis: A Dataset for Enhancing {MLLM}s in Visualization Understanding},
  author = {Liu, Can and Da, Chunlin and Long, Xiaoxiao and Yang, Yuxiao and Zhang, Yu and Wang, Yong},
  booktitle = {2025 IEEE Visualization and Visual Analytics (VIS)},
  year = {2025},
  pages = {26--30},
  publisher = {IEEE},
  doi = {10.1109/VIS60296.2025.00010}
}

@inproceedings{ahn2025chatgpt,
  title = {Understanding Why {ChatGPT} Outperforms Humans in Visualization Design Advice},
  author = {Ahn, Yongsu and Kim, Nam Wook},
  booktitle = {2025 IEEE Visualization and Visual Analytics (VIS)},
  year = {2025},
  pages = {166--170},
  publisher = {IEEE},
  doi = {10.1109/VIS60296.2025.00039}
}

@inproceedings{xaireverse2025,
  title = {Enhancing {XAI} Interpretation through a Reverse Mapping from Insights to Visualizations},
  author = {Nuthalapati, Aniket and Hinds, Nicholas and Lim, Brian Y. and Wang, Qianwen},
  booktitle = {2025 IEEE Visualization and Visual Analytics (VIS)},
  year = {2025},
  pages = {41--45},
  publisher = {IEEE},
  doi = {10.1109/VIS60296.2025.00013}
}

@inproceedings{bnnvis2025,
  title = {{BNNVis}: Towards Visual Analytics for Bayesian Neural Networks},
  author = {Appleby, Gabriel and Hassanaly, Malik and Rogers, Jen and Mueller, Juliane and Potter, Kristi},
  booktitle = {2025 IEEE Visualization and Visual Analytics (VIS)},
  year = {2025},
  pages = {146--150},
  publisher = {IEEE},
  doi = {10.1109/VIS60296.2025.00035}
}

@inproceedings{revise2025,
  title = {{ReVise}: A Human-{AI} Interface for Incremental Algorithmic Recourse},
  author = {Bhattacharjee, Kaustav and Yuan, Jun and Dasgupta, Aritra},
  booktitle = {2025 IEEE Visualization and Visual Analytics (VIS)},
  year = {2025},
  pages = {151--155},
  publisher = {IEEE},
  doi = {10.1109/VIS60296.2025.00036}
}

@misc{reflection2025,
  title = {Reflection on Data Storytelling Tools in the Generative {AI} Era from the Human-{AI} Collaboration Perspective},
  author = {Li, Haotian and Wang, Yun and Qu, Huamin},
  year = {2025},
  eprint = {2503.02631},
  archivePrefix = {arXiv},
  primaryClass = {cs.HC},
  doi = {10.48550/arXiv.2503.02631},
  url = {https://arxiv.org/abs/2503.02631}
}

@inproceedings{logicdecisionaid2025,
  title = {Toward a Logic of Generalization about Visualization as a Decision Aid},
  author = {Kale, Alex},
  booktitle = {2025 IEEE Visualization and Visual Analytics (VIS)},
  year = {2025},
  pages = {1--5},
  publisher = {IEEE},
  doi = {10.1109/VIS60296.2025.00005}
}

@inproceedings{wikimixqa2025,
  title = {WikiMixQA: A Multimodal Benchmark for Question Answering over Tables and Charts},
  author = {Foroutan, Negar and Romanou, Angelika and Ansaripour, Matin and Eisenschlos, Julian Martin and Aberer, Karl and Lebret, R{\'e}mi},
  booktitle = {Findings of the Association for Computational Linguistics: ACL 2025},
  year = {2025},
  pages = {24941--24958},
  address = {Vienna, Austria},
  publisher = {Association for Computational Linguistics}
}

@misc{finsage2025,
  title = {FinSage: A Multi-aspect {RAG} System for Financial Filings Question Answering},
  author = {Wang, Xinyu and Chi, Jijun and Tai, Zhenghan and others},
  year = {2025},
  eprint = {2504.14493},
  archivePrefix = {arXiv},
  primaryClass = {cs.CL},
  doi = {10.48550/arXiv.2504.14493},
  url = {https://arxiv.org/abs/2504.14493}
}

@misc{multifinrag2025,
  title = {MultiFinRAG: An Optimized Multimodal Retrieval-Augmented Generation ({RAG}) Framework for Financial Question Answering},
  author = {Gondhalekar, Chinmay and Patel, Urjitkumar and Yeh, Fang-Chun},
  year = {2025},
  eprint = {2506.20821},
  archivePrefix = {arXiv},
  primaryClass = {cs.CL},
  doi = {10.48550/arXiv.2506.20821},
  url = {https://arxiv.org/abs/2506.20821}
}

@misc{multifinben2025,
  title = {MultiFinBen: A Multilingual, Multimodal, and Difficulty-Aware Benchmark for Financial {LLM} Evaluation},
  author = {Peng, Xueqing and Qian, Lingfei and Wang, Yan and others},
  year = {2025},
  eprint = {2506.14028},
  archivePrefix = {arXiv},
  primaryClass = {cs.CL},
  doi = {10.48550/arXiv.2506.14028},
  url = {https://arxiv.org/abs/2506.14028}
}

@misc{aida2025layout,
  title = {Enhancing Large Vision-Language Models with Layout Modality for Table Question Answering on Japanese Annual Securities Reports},
  author = {Aida, Hayato and Takahashi, Kosuke and Omi, Takahiro},
  year = {2025},
  eprint = {2505.17625},
  archivePrefix = {arXiv},
  primaryClass = {cs.CL},
  doi = {10.48550/arXiv.2505.17625},
  url = {https://arxiv.org/abs/2505.17625}
}

\end{document}